\documentclass[%
reprint,
superscriptaddress,
nofootinbib,
amsmath,amssymb,
aps,
prd,
floatfix,
]{revtex4-1}

\usepackage[usenames,dvipsnames]{color}
\usepackage{graphicx}
\usepackage{dcolumn}
\usepackage{bm}
\usepackage{hyperref}
\usepackage{float} 
\usepackage[normalem]{ulem}

\hypersetup{
    colorlinks=true,
    linkcolor=blue,
    citecolor=blue,
    filecolor=black,
    urlcolor=blue,
}

\def\calP{{\cal P}}

\def\calN{{\cal N}}

\def\|{{ \, || \,}}

\newcommand{\fnl}{f_{\rm NL}}
\def\g{{\gamma}}
\def\G{{\Gamma}}
\def\tosc{{t^{\rm osc}_i}}

\newcommand{\be}{\begin{equation}}
\newcommand{\ee}{\end{equation}}
\newcommand{\bea}{\begin{eqnarray}}
\newcommand{\eea}{\end{eqnarray}}

\begin{document}

\begin{center}
\rightline{\small DESY-17-169}
\vskip -3cm
\end{center}

\title{Effect of reheating on predictions following multiple-field inflation}

\author{Selim~C.~Hotinli}
	\email{selim.hotinli14@imperial.ac.uk}
	\affiliation{Astrophysics Group \& Imperial Centre for Inference and Cosmology, Department of Physics, Imperial College London, Blackett Laboratory, Prince Consort
Road, London SW7 2AZ, UK}

\author{Jonathan~Frazer}
  \email{jonathan.frazer@desy.de}
  \affiliation{Deutsches Elektronen-Synchrotron DESY, Theory Group, 22603 Hamburg, Germany}

\author{Andrew~H.~Jaffe}
	\email{a.jaffe@imperial.ac.uk}
    \affiliation{Astrophysics Group \& Imperial Centre for Inference and Cosmology, Department of Physics, Imperial College London, Blackett Laboratory, Prince Consort
Road, London SW7 2AZ, UK}

\author{Joel~Meyers}
  \email{jmeyers@cita.utoronto.ca}
  \affiliation{Canadian Institute for Theoretical Astrophysics, University of Toronto, Toronto, Ontario M5S 3H8, Canada}

\author{Layne~C.~Price}
  \email{laynep@andrew.cmu.edu}
  \affiliation{McWilliams Center for Cosmology, Department of Physics, Carnegie Mellon University, Pittsburgh, PA 15213, USA}

\author{Ewan~R.M.~Tarrant}
  \email{e.tarrant@sussex.ac.uk}
  \affiliation{Astronomy Centre, University of Sussex, Falmer, Brighton BN1 9QH, UK}

\date{\today}

\begin{abstract}

We study the sensitivity of cosmological observables to the reheating phase following inflation driven by many scalar fields. We describe a method which allows semi-analytic treatment of the impact of perturbative reheating on cosmological perturbations using the sudden decay approximation.  Focusing on $\mathcal{N}$-quadratic inflation, we show how the scalar spectral index and tensor-to-scalar ratio are affected by the rates at which the scalar fields decay into radiation.  We find that for certain choices of decay rates, reheating following multiple-field inflation can have a significant impact on the prediction of cosmological observables.

\end{abstract}

\maketitle



\section{Introduction}

The inflationary paradigm~\cite{Starobinsky:1979ty,Starobinsky:1980te,Kazanas:1980tx,Sato:1980yn,Guth:1980zm} solves many of the classical problems associated with the hot Big Bang scenario, while providing a natural mechanism for generating primordial cosmological fluctuations~\cite{Mukhanov:1981xt,Hawking:1982cz,Guth:1982ec,Starobinsky:1982ee,Bardeen:1983qw}. Observations are currently consistent with the simplest single-field, slow-roll models of inflation, \emph{e.g.}, the \emph{Planck} observations of the cosmic microwave background (CMB)~\cite{Ade:2015lrj} indicate a featureless power-law shape for the primordial power spectrum of scalar fluctuations and no detectable primordial non-Gaussianity or tensor fluctuations.

The predictions of single-field inflation are largely insensitive to the details of reheating.  Single-field inflation models produce purely adiabatic curvature perturbations $\zeta_k$, which guarantees that the $n$-point correlation functions, $\langle\zeta^n\rangle$, do not evolve on scales exceeding the Hubble radius $k \lesssim aH$ during and after inflation~\cite{Weinberg:2008zzc,Weinberg:2003sw,Weinberg:2004kr,Weinberg:2008nf}.\footnote{We mean by `adiabatic fluctuations' those for which the perturbation to any four-scalar in the system is proportional to the rate of change of the scalar, with the same proportionality for all scalars.} As a consequence, post-inflationary dynamics will not cause the perturbation spectra to evolve on super-Hubble scales, and only the integrated expansion following single-field inflation affects the prediction of cosmological observables~\cite{Dodelson:2003vq,Liddle:2003as,Alabidi:2005qi,Adshead:2010mc,Easther:2011yq,Martin:2014nya}.

Despite the phenomenological success of single-field models, they lack the generality of more complex scenarios, representing only a limited class of possible models. Importantly, they are not always natural from a theoretical point of view, \emph{e.g.}, string compactifications often result in hundreds of scalar fields appearing in the low energy effective action~\cite{Grana:2005jc,Douglas:2006es,Denef:2007pq,Denef:2008wq}. Models with multiple fields naturally produce non-adiabatic fluctuations, whose presence allows the curvature perturbation and its correlation functions to evolve outside the Hubble radius.  Therefore, in order to make predictions in multi-field models, it is necessary to understand the evolution of the correlation functions until either the curvature fluctuations become adiabatic or they are directly observed. Unless an `adiabatic limit' \cite{Meyers:2010rg,Meyers:2011mm,Seery:2012vj,Elliston:2011dr,Polarski:1994rz,GarciaBellido:1995qq,Langlois:1999dw} is established before the onset of reheating, then the observable predictions of multi-field models will be sensitive to post-inflationary dynamics that must be accurately modeled before comparing the results to data.

Non-adiabatic fluctuations can become adiabatic if the Universe passes through a phase of effectively single-field inflation~\cite{Meyers:2010rg,Meyers:2011mm,Seery:2012vj,Elliston:2011dr,Polarski:1994rz,GarciaBellido:1995qq,Langlois:1999dw} or through a period of local thermal and chemical equilibrium with no non-zero conserved quantum numbers~\cite{Weinberg:2004kf,Weinberg:2008si,Meyers:2012ni}. The latter conditions are often established during the late stages of reheating, though notable exceptions include models in which dark matter is not a thermal relic, or where baryon number was produced before the end of inflation~\cite{Bucher:1999re}.  We will assume throughout this work that the result of reheating is a relativistic thermal plasma described entirely by its temperature.  In this paper we will focus on developing a methodology for calculating the predictions of multi-field inflation for the fully adiabatic power spectrum of curvature perturbations after reheating.

For two-field inflation, numerical studies~\cite{Elliston:2011dr,Leung:2012ve,Leung:2013rza} have demonstrated that observables such as the power spectrum $\calP(k)$, and the local shape bispectrum parameter $\fnl$ can be very sensitive to the details of reheating.  This sensitivity was quantified in Ref.~\cite{Meyers:2013gua}, where it was shown that the adiabatic observables take values within finite ranges that are determined completely by the details of the underlying inflationary model.
The effect of reheating is to preferentially enhance or suppress the initial fluctuations of some fields compared to others, depending on the details of the reheating model.  This gives predictions that effectively interpolate between those obtained by projecting the non-adiabatic perturbations along each of the two field direction $\phi_i$ in isolation.
If the projection into each direction is the same, then the sensitivity to reheating for two-field inflation models is minimal.

In this paper we extend the results of Ref.~\cite{Meyers:2013gua} and provide a general methodology for calculating the adiabatic power spectrum of curvature perturbations after multi-field inflation for any number of scalar fields. The regime of many-field inflation ($\calN \gtrsim 10$) typically yields a range of predictions for curvature perturbations at the end of inflation that is surprisingly easy to categorize in comparison to the apparently large dimensionality of parameter space (see \emph{e.g.} Refs~\cite{Frazer:2013zoa, Kaiser:2013sna,Easther:2013rva,Wenren:2014cga,Price:2014ufa,Price:2015qqb,Dias:2016slx,Dias:2017gva,Bjorkmo:2017nzd}, though stochastic effects can be important in the presence of many fields~\cite{Vennin:2016wnk}).
Scenarios with many fields also tend to predict an amount of isocurvature perturbations at the end of inflation which increases with the number of fields~\cite{Easther:2013rva, Dias:2017gva}, thereby elevating the importance of studying the effects of reheating for these models.

As in Ref.~\cite{Meyers:2013gua}, we restrict ourselves to perturbative reheating.  This ignores interesting dynamics such as preheating, which may non-perturbatively produce radiation quanta through parametric resonance~\cite{Greene:1997fu} potentially leading to rich phenomena including primordial non-Gaussianity \cite{enqvist2005non,Bond:2009xx,Kohri:2009ac} and perhaps the production of primordial black holes \cite{Green:2000he,Bassett:2000ha,khlopov2010primordial}.
However, perturbative reheating is a generically good phenomenological description for inflationary models with many degrees of freedom, as periods of exponential particle production become much harder to realize when many fields must conspire together to resonate~\cite{Battefeld:2008bu,Battefeld:2009xw,Braden:2010wd}, although single-field attractor behavior is common for some multiple-field models with non-minimal couplings to gravity~\cite{2015arXiv151008553D}.
Therefore, the methodology we develop here is quite generic for inflation with many fields $\calN \gg 2$.


\section{Overview}\label{sec:overview}

We begin here with a broad description of the methods that will be described in more detail in subsequent sections.  We are interested in calculating the two-point statistics of the curvature perturbation after reheating has completed following multiple-field inflation.   
We will focus in particular on the scalar spectral index $n_s$ and the tensor-to-scalar ratio $r$.

The $\delta N$ formalism is a useful method for calculating the superhorizon evolution of the curvature perturbation in terms of the initial fluctuations of a set of scalar fields~\cite{Salopek:1990jq,Sasaki:1995aw,Sasaki:1998ug,Wands:2000dp,Lyth:2004gb}. In this method one calculates the expansion from some initial time $t_\star$ on a spatially-flat hypersurface $g_{ij}(t_\star,\textbf{x})=a^2(t_\star)\delta_{ij}$, to some final time $t_c$ on a uniform density hypersurface $\rho(t_c,\mathbf{x}) = \bar{\rho}(t_c)$.  In practice we will take the initial hypersurface to be at horizon exit and the uniform density hypersurface to be after the conclusion of reheating when the Universe is dominated by a thermal bath of radiation.   The number of $e$-folds of expansion, defined as $N=\ln{a_c}/{a_\star}$, is given by
\be 	
	N(t_\star,t_c) = \int_{t_\star}^{t_c} H(t)\mathrm{d}t \, .
\ee
The perturbation to the number of $e$-foldings of expansion is equal to the difference in the curvature perturbation on these two hypersurfaces
\be\label{eq:delta_N}
	\zeta = \delta N = \sum_i N_{,i}\delta\phi_i^\star +\frac12 \sum_{ij}N_{,ij}\delta\phi_i^\star\delta\phi_j^\star+\ldots \, ,
\ee 
where $N_{,i}$ refers to the derivative of the number of $e$-folds of expansion with respect to the initial scalar field value $N_{,i} =\partial N/\partial \phi_i^\star$.

Using Eq.~\eqref{eq:delta_N} we can then calculate the observables of interest.  Focusing on the two-point statistics, we find the curvature power spectrum,
\be \label{eq:curv_power}
\mathcal{P}_\zeta=\mathcal{P}_\star\sum\limits_iN^2_{,i}\ ,
\ee 
the scalar spectral index,
\be 
n_s-1=-2\epsilon_\star-\frac{2}{\sum_iN^2_{,i}}\left[1-\sum_{ij}\eta_{ij}^\star N_{,i}N_{,j}\right]\ , 
\ee
and the tensor-to-scalar ratio
\be 
r = \frac{8P_\star}{P_\zeta}\ , 
\ee 
where sums are carried out over all field indices ${i=1\ldots{\mathcal{N}}}$. We have introduced the initial spectrum of scalar field fluctuations $\mathcal{P}_\star=H^2_\star/2k_\star^3$, and the slow-roll parameters $\epsilon_\star=-(\dot{H}/H^2)_\star$ and $\eta^\star_{ij}=(V_{,ij}/V)_\star$, which are calculated at horizon crossing. 

In order to calculate the expansion history and how it depends on the initial scalar field configuration, one in general needs to solve the perturbed field equations from horizon exit all the way through reheating.  This is typically quite challenging due to the wide range of time and energy scales involved in the problem.  The methods we will describe allow us to treat the post-inflationary evolution in a simplified manner, thus greatly reducing the computational cost of making predictions in multi-field inflationary models.

We proceed by splitting the problem into two parts.  We first treat the evolution from horizon exit through inflation to a phase where the scalar fields are coherently oscillating about the minima of their potentials.  This portion of the evolution is treated by numerically solving the perturbed field equations and is described in detail in Sec.~\ref{sec:inf_perturbations}.  Next, we treat the process of reheating, when the scalar fields decay into radiation.  As described in detail in Sec.~\ref{sec:sudden_dec}, this part of the evolution can be treated semi-analytically by using a fluid approximation at very low computational cost, thus allowing us to quickly calculate how a wide range of reheating scenarios impacts the observable predictions of a particular multi-field inflationary model. For this part of the evolution, the unperturbed fluid equations are evaluated numerically, and the sudden decay approximation is applied to determine the impact of reheating on the cosmological perturbations.

As will be shown below, the impact of reheating following multiple field inflation is to mix together perturbations present in individual scalar fields present at the end of inflation into the final curvature perturbation with weights determined by the reheating parameters. Additionally, reheating impacts how the length scales we observe today are related to the scales during inflation. Even in single-field inflation, reheating affects how many $e$-foldings $N_\star$ before the end of inflation the observed fluctuations have crossed the Hubble horizon. 
Predicting this quantity requires matching the Hubble scale today to the Hubble scale during inflation, hence the modeling of the entire expansion history of the Universe. A simple comparison (approximating transitions between different epochs in the history of the Universe as instantaneous and ignoring the recent phase of dark energy domination) can be made by using the classical matching equation~\cite{Liddle:1993fq,Liddle:2003as} 
\be\label{eq:matching_efolds}
\frac{k_\star}{a_0H_0}
=e^{-N_\star}\frac{a_\text{end}}{a_\text{reh}}\frac{a_\text{reh}}{a_\text{eq}}\frac{H_\star}{H_\text{eq}}\frac{a_\text{eq}H_\text{eq}}{a_0H_0}\ ,
\ee
where the number of $e$-foldings between the end of inflation and when the pivot scale crosses the Hubble horizon $k_\star=a_\star H_\star$ is defined as ${N_\star}=\ln a_\text{end}/a_\star$, and $a_\text{reh}$ is the scale factor at the the end of reheating, i.e. after all fields have decayed into radiation. The remaining quantities in the above expression are the Hubble horizon $H_0$ and the scale factor $a_0$ today and at the time of matter-radiation equality: $H_\text{eq}$, $a_\text{eq}$. The latter four quantities are well known from large-scale observations of the Universe. The remaining quantities are predicted by the inflationary model and the details of reheating, which fixes $N_\star$, the number of $e$-foldings of inflation after the pivot scale exits the horizon.


\section{Reheating}\label{sec:sudden_dec}

Regardless of the inflationary model, or how many scalar fields were present during inflation, the universe must eventually evolve to the radiation-dominated era of the standard Big Bang model. This can be achieved by coupling the fields $\phi_i$ to relativistic particle species. As the fields approach, overshoot, and begin to oscillate about the minimum of their potentials, interactions with lighter particles lead to dissipation which drains energy from the $\phi_i$ zero-mode and excites relativistic particles. We refer to these collective processes as reheating (see \emph{e.g.},~\cite{Kofman:1997yn,Allahverdi:2010xz, amin2015nonperturbative} for reviews).

The relativistic energy densities gain energy at a rate
\be
  \dot \rho_i^\gamma + 4H \rho_i^\gamma = \Gamma_i \rho_i
\label{eq:gammaEoM}
\ee
whilst damping of the inflaton zero mode due to this decay process can be approximated by 
\be
\ddot\phi_i+(3H+\G_{i})\dot\phi_i+m_i^2\phi_i=0\,,
\label{eq:phiEoM}
\ee
and the energy density stored in the oscillating field is $\rho_i=\frac12(\dot\phi_i^2+m_i^2\phi_i^2)$. Perturbative decay of the oscillating fields relies on the assumption that the decay rates can be calculated by standard methods in perturbative quantum field theory. If, however, the amplitude of the field oscillations, and the couplings to gauge fields are sufficiently large, perturbation theory breaks down and reheating proceeds in a different way, through parametric resonance~\cite{Shtanov:1994ce,Kofman:1994rk,Traschen:1990sw}. 

The impact of reheating on cosmological observables is well captured by appealing to the sudden decay approximation~\cite{Lyth:2001nq,Lyth:2005fi,Sasaki:2006kq}. This approximation has been used frequently in the past to calculate the statistics of the primordial curvature perturbation for various models of inflation~\cite{Lyth:2005fi,Sasaki:2006kq,Kawasaki:2011pd,Kawasaki:2012gg,Fonseca:2012cj}, the most widely known example with multiple fields being the curvaton scenario~\cite{Linde:1996gt,Enqvist:2001zp,Lyth:2001nq,Lyth:2002my,Vennin:2015vfa}. Furthermore, numerical studies have shown that for the curvaton scenario, and general models of two--field inflation, sudden decay reproduces the gradual decay result obtained by solving Eqs.~(\ref{eq:gammaEoM}) and ~(\ref{eq:phiEoM}) (together with the Friedmann constraint) remarkably well~\cite{Sasaki:2006kq,Meyers:2013gua}.

\begin{figure}[!t]
\includegraphics[width=8.6cm]{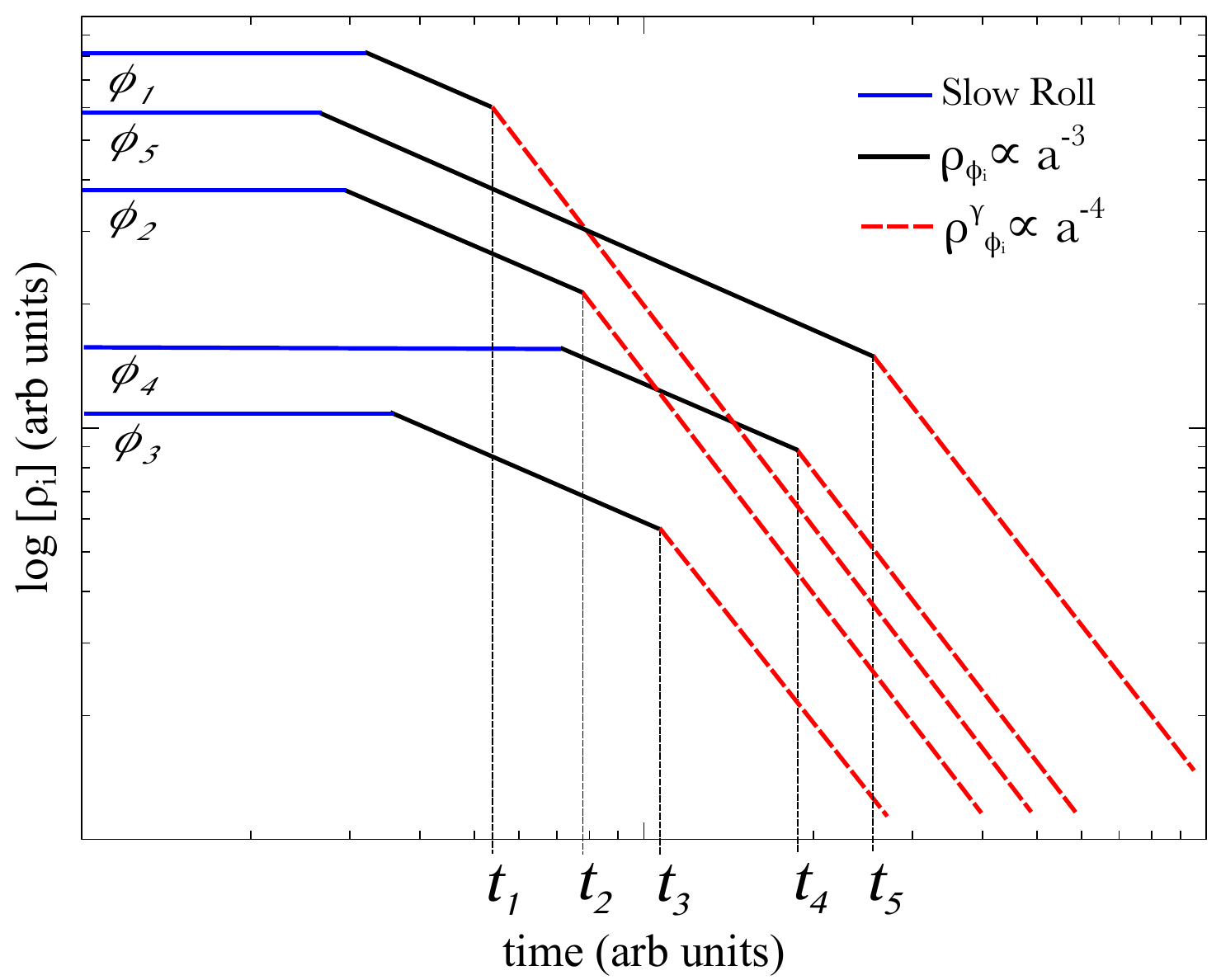}
\caption{A typical ``sudden decay'' energy diagram illustrating the decay of $\calN=5$ fields. After the fields leave slow roll (blue sold lines) they begin to oscillate about their quadratic minima, their energy density scaling as $a^{-3}$ (back sold lines). When $H(t_i)=\G_i$ field $\phi_i$ decays instantly into radiation (red dashed lines) which scales as $a^{-4}$.}
\label{fig:decay_schematic}
\end{figure}

We will focus on fields $\phi_i$ rolling in potentials with quadratic minima.  During the phase of coherent oscillations, we will treat these fields as perfect fluids with vanishing pressure.  In this approximation, the density of these matter fluids scale as $a^{-3}$ and do not interact with their decay products until they instantly decay at some time $t_i$. These dynamics are schematically illustrated in Fig.~\ref{fig:decay_schematic} for the specific example of $\calN=5$ fields. We are interested in the statistics of the curvature perturbation $\zeta(t,{\bf x})$ at the final time $t=t_f$, when reheating has completed (all fields have decayed). Field $\phi_i$ (represented in Fig.~\ref{fig:decay_schematic} by its energy density $\rho_i$) is labelled according to its decay time $t_i$, where $i=1,2,\ldots,\calN$ and $t_1<t_2<\ldots<t_\calN$. With this notation, the final time $t_f=t_\calN$.  Our derivation in this section is a generalization of the methods described in Refs.~\cite{Sasaki:2006kq,Assadullahi:2007uw,Meyers:2013gua}.

The underlying assumption of the sudden decay approximation is that the fields decay instantly into radiation when the Hubble rate becomes equal to the decay rate $H(t_i)=\G_{i}$, which defines the decay time $t_i$. Furthermore, the decay hypersurfaces are taken to be surfaces of uniform energy density, upon which
\bea
\bar\rho_{\rm tot}(t_1) &=& \bar\rho_\phi(t_1)=\sum_{i=1}^\calN \rho_i(t_1,{\bf x})\,, \label{eq:decay-surface-0} \\
\bar\rho_{\rm tot}(t_j) &=& \rho_\g(t_j,{\bf x}) + \rho_\phi(t_j,{\bf x}) \nonumber \\ 
&=& \sum_{i=1}^{j-1} \rho^\g_i(t_j,{\bf x}) + \sum_{i=j}^{\calN} \rho_i(t_j,{\bf x})\,, \quad j\ge2\,,
\label{eq:decay-surface}
\eea
where $\rho_i(t_j,{\bf x}) = \bar\rho_i(t_j) + \delta\rho_i(t_j,{\bf x})$ and $\rho^\g_i(t_j,{\bf x}) = \bar\rho^\g_i(t_j) + \delta\rho^\g_i(t_j,{\bf x})$. Here, $\rho_\phi$ denotes the total energy density stored in the oscillating scalar fields, and $\rho_\g$ denotes the total energy density stored in the decay products
\bea
\rho_\phi = \sum_i \rho_i \, , \qquad \rho_\g = \sum_i \rho_i^\g \, .
\eea

Our first task is to determine how the individual curvature perturbation, $\zeta_i$, associated with field $\phi_i$, passes its fluctuation over to its decay product, $\zeta^\g_i$. Within the confines of the sudden decay approximation this conversion is instantaneous.  In the absence of interactions, fluids with barotropic equation of state, such as dust--like oscillating scalar fields and their radiation fluid decay products, have an individually conserved curvature perturbation~\cite{Lyth:2004gb,Sasaki:2006kq}
\be
\zeta_i = \delta N +\frac13\int^{\rho_i(t,{\bf x})}_{\bar\rho_i(t)} 
\frac{{\rm d}\tilde\rho_i}{\tilde\rho_i + P_i(\tilde\rho_i)}\,.
\label{eq:SD:zeta_i}
\ee
Here, $\delta N$ is the perturbed amount of expansion, which working within the separate universe assumption~\cite{Wands:2000dp,Lyth:2004gb}, is equivalent to the difference in curvature perturbations measured from an initial flat hypersurface, up to one of constant energy density: $\delta N=\zeta$. In this notation,  fluctuations are purely adiabatic if $\zeta_i=\zeta$ for all constituents of the Universe.


From this point on, all unbarred quantities will have an implicit dependence on position, while barred quantities have no spatial dependence. With $P_i=0$ (relevant for the dust--like oscillating scalar fields before they decay), and $P_i^\g=\rho_i^\g/3$ (for the radiation decay products) we can easily perform the integral in Eq.~(\ref{eq:SD:zeta_i}) to find:
\bea
\rho_i(t_j) &=& \bar\rho_i(t_j) e^{3(\zeta_1(t_1)-\zeta(t_1))}\,, \quad i\ge j\,, \label{eq:matching-0} \\ 
\rho^\g_i(t_j) &=& \bar\rho^\g_i(t_j) e^{4(\zeta^\g_1(t_1)-\zeta(t_1))}\,, \quad i\le j\,.
\label{eq:matching-1}
\eea
The $i\ge j$ and $i\le j$ conditions reflect the fact that the decay products do not exist until the field has decayed. We have retained the explicit $t_j$ dependence for the individual $\zeta_i$ for clarity, but it is to be understood that $\zeta_i$ is conserved between $\tosc\le t\le t_i$, where $\tosc$ is the time then the field $\phi_i$ begins to oscillate. 

Making use of Eqs.~(\ref{eq:decay-surface-0}) and~(\ref{eq:matching-0}), we have on the first decay hypersurface:
\be
1 = \sum_{i=1}^\calN\Omega_i(t_1)e^{3(\zeta_1(t_1)-\zeta(t_1))}\,, \quad 
\Omega_i(t_1) \equiv \frac{\bar\rho_i}{\bar\rho_\phi}\Bigg|_{t_1}\,.
\label{eq:zeta1-nonlin}
\ee
Since decay is instantaneous, $\rho_1^\g(t_1,{\bf x}) = \rho_1(t_1,{\bf x})$, which making use of Eq.~(\ref{eq:matching-1}), is equivalent to
\be
\bar\rho_1^\g(t_1) e^{4(\zeta_1^\g(t_1)-\zeta(t_1))} = \bar\rho_1(t_1) e^{3(\zeta_1(t_1)-\zeta(t_1))}\,.
\ee
This must hold true even in the absence of fluctuations (where $\bar\rho_1^\g(t_1) =\bar\rho_1(t_1)$) and so
\be
\zeta_1^\g(t_1) = \frac34 \zeta_1(t_1) + \frac14 \zeta(t_1)\,.
\ee
This expression provides the matching condition for the curvature perturbation on surfaces of uniform $\rho_1$ and uniform $\rho_1^\g$ either side of the decay time $t_1$, and straightforwardly generalises to all subsequent decay times:
\be
\zeta_i^\g(t_i) = \frac34 \zeta_i(t_i) + \frac14 \zeta(t_i)\,.
\label{eq:matching}
\ee

Having determined these matching conditions, we seek an expression for the total curvature perturbation at time $t_\calN$. This is straightforward to obtain by repeating the above calculation for all subsequent decay times. Using Eqs.~(\ref{eq:matching-1}) and~(\ref{eq:decay-surface}), we find, for $j\ge2$:
\be
1 = \sum_{i=1}^{j-1}\Omega_i^\g(t_j)e^{4(\zeta_i^\g(t_j)-\zeta(t_j))} + \sum^{\calN}_{i=j}\Omega_i(t_j)e^{3(\zeta_i(t_j)-\zeta(t_j))}\,,
\label{eq:zetatN-nonlin}
\ee
where
\be
\Omega_i(t_j)=\frac{\bar\rho_i}{\bar\rho_\g+\bar\rho_\phi}\Bigg|_{t_j}\,, \qquad   \Omega^\g_i(t_j)=\frac{\bar\rho^\g_i}{\bar\rho_\g+\bar\rho_\phi}\Bigg|_{t_j}\,.
\ee
Eq.~(\ref{eq:zetatN-nonlin}) constitutes a non--linear expression for $\zeta(t_\calN,{\bf x})$ if one takes $j=\calN$. In order to solve Eq.~(\ref{eq:zetatN-nonlin}) for $\zeta(t_\calN,{\bf x})$, we proceed perturbatively. Expanding to first order and rearranging slightly:
\be
\zeta(t_j) = 
\frac43\sum_{i=1}^{j-1}r_{ij}\zeta_i^\g(t_j) + \sum_{i=j}^\calN r_{ij}\zeta_i(t_j)\,, \quad j\ge2\,,
\label{eq:zetaN-final-1}
\ee
where we have defined the `sudden decay parameters'
\be
r_{ij} \equiv r_i(t_j) \equiv
\begin{cases}
 \frac{3\bar\rho_i}{4\bar\rho_\g+3\bar\rho_\phi}\Big|_{t_{j}} \quad \text{for} \quad i\geq j \\
 \frac{3\bar\rho^\g_i}{4\bar\rho_\g+3\bar\rho_\phi}\Big|_{t_{j}} \quad \text{for} \quad i< j \,.
  \end{cases}
\label{eq:rij}
\ee
Since the $\zeta^\g_i$ are conserved for $t\ge t_i$, we may write $\zeta_i^\g(t_j)=\zeta_i^\g(t_i)$ for $i\le j$, and use Eq.~(\ref{eq:matching}) in Eq.~(\ref{eq:zetaN-final-1}) to substitute for $\zeta_i^\g(t_i)$. Similarly, we may write $\zeta_i(t_j)=\zeta_i(\tosc)$ for $i\ge j$. Making these two replacements, we find:
\be
\zeta(t_j) = \frac13\sum_{i=1}^{j-1}r_{ij}\zeta(t_i) + \sum_{i=1}^\calN r_{ij}\zeta_i(\tosc) \, .
\ee
Evaluating this expression for $j=\calN$ gives us a recursive expression for curvature perturbation at the end of reheating.  After some straightforward manipulation, the expression for $\zeta(t_\calN)$ can be put into a slightly more convenient form:
\be
\zeta(t_\calN,{\bf x}) = \sum_{i=1}^\calN W_i\, \zeta_i(\tosc,{\bf x})\,,
\label{eq:zetaN1}
\ee
where
\be 
W_i=\sum_{j=0}^{\calN-1} A_j\,  r_{i(\calN-j)} \, ,
\label{eq:weights}
\ee
and we have defined
\be 
A_j=\frac{1}{3}\sum_{k=0}^{j-1} A_k \, r_{(\calN-j)(\calN-k)}  \, ,
\label{eq:weight_coef}
\ee
and $A_0=1$. 
Eq.~(\ref{eq:zetaN1}) is our final expression for the primordial curvature perturbation at the completion of reheating. It is the statistics $\langle\zeta^n(t_\calN,{\bf x})\rangle$ of this fluctuation that are relevant for observation. It is clear from Eq.~(\ref{eq:zetaN1}) that the effect of reheating (captured by the weights $W_i$) is to re--scale the $\zeta_i(\tosc,{\bf x})$. The $W_i$ are functions of the sudden decay parameters $r_{ij}$, which can be directly related to the physical decay rates $\G_i$ within the confines of the sudden decay approximation. As discussed in~\cite{Meyers:2013gua}, this is one area where the sudden decay approximation falls short and for this reason we compute the mapping from $\G_i$ to $r_{ij}$  numerically. 

The individual curvature fluctuations $\zeta_i(\tosc,{\bf x})$ are determined completely by the details of inflation (the form of the potential and the field values at horizon crossing), and do not depend in any way upon reheating. As can be seen from Eqs.~(\ref{eq:zetaN1})-(\ref{eq:weight_coef}) once the curvature fluctuations are known, the effect of reheating on the cosmological perturbations can be calculated using only unperturbed energy densities evaluated at various times during the reheating phase. In the following section we discuss the calculation of the curvature fluctuations resulting from inflation.


\section{Inflationary Perturbations}\label{sec:inf_perturbations}

Generically, $\zeta_i(t,{\bf x})$ will evolve during multi-field inflation until an adiabatic limit is reached, at which point they become equal and conserved~\cite{Meyers:2010rg,Meyers:2011mm,Elliston:2011dr}. Whether conservation is achieved before the end of inflation depends upon the specifics of the inflationary model. Regardless of these specifics however, it is guaranteed that the $\zeta_i(t,{\bf x})$ will (to a very good approximation) be conserved quantities during the period when field $\phi_i$ is oscillating and before it has decayed appreciably into radiation. It is therefore sufficient to compute these quantities at $t=\tosc$. 

We use the publicly available \textsc{MultiModeCode} inflation solver~\cite{Mortonson:2010er,Easther:2011yq,Norena:2012rs,Price:2014xpa} to evaluate the first-order mode equations for each scalar field, without using the slow-roll approximation.  Following the convention of Ref.~\cite{Salopek:1988qh} we expand each of the first-order field perturbations in terms of a complex valued matrix $q_{ij}$ as
\begin{equation}\label{eq:modematrix}
  {\delta \phi^i} (t, \vec k) = q_{ij} (t, k) \hat a^j (\vec k) + q^*_{ij}(t, k) \hat{a}^{\dagger,j}(- \vec k)\ ,
\end{equation}
where the creation and annihilation operators satisfy $(\hat a^j(\vec k))^\dagger = \hat{a}^{\dagger,j}(- \vec k) $.
The transformed variable $\psi_{ij} = q_{ij}/a$ satisfies the Mukhanov-Sasaki equation of motion with a ``mixed'' mass matrix $\mathcal M_{ij}$
\be
\frac{{\rm d}^2\psi_{ij}}{{\rm d} N^2}+(1-\epsilon)\frac{{\rm d}\psi_{ij}}{{\rm d} N}+\left(\frac{k^2}{a^2H^2}-2-\epsilon\right)\psi_{ij}+\mathcal{M}_{im}\psi^{m}_{j}=0
\label{eqn:kg_for_muksas}
\ee
where
\be 
\begin{split}
\mathcal{M}_{ij}&\equiv\frac{\partial_i\partial_jV}{H^2}+\frac{1}{H^2}\left(\frac{{\rm d}\phi_i}{{\rm d}N}\partial_jV+\frac{{\rm d}\phi_j}{{\rm d}N}\partial_iV\right)\\ 
&+(3-\epsilon)\frac{{\rm d}\phi_i}{{\rm d}N}\frac{{\rm d}\phi_j}{{\rm d}N}\ , 
\end{split}
\ee
with $\partial_i\equiv\partial/\partial\phi_i$ and $N$ is the number of $e$-folds. We use the Bunch-Davies initial condition~\cite{Bunch:1978yq} for the transformed variable $\psi_{ij} \sim \delta_{ij}$. 

The components of curvature perturbation are defined in the spatially-flat gauge as 
\be\label{eq:zeta_spatfloat}
\zeta_i(t,\vec{k})\equiv\frac{H}{\dot{\bar{\rho}}_i}\delta\rho_i(t,\vec{k})\ . 
\ee
The density perturbations $\delta\rho_i(t,\vec{k})$ are given by 
\be\label{eq:density_pert}
\begin{split}
\delta\rho_i(t,\vec{k})&=\dot{\phi}_i(t)\dot{\delta\phi}_i(t,\vec{k}) \\ 
&-\frac{\dot{\phi}_i(t)^2}{2H}\sum\limits_m\dot{\phi}_m(t)\delta\phi_m(t,\vec{k})+V_{,i}\delta\phi_i(t,\vec{k})\ . 
\end{split}
\ee 

Similar to the field perturbations, we expand each of the curvature perturbation components in the same basis by defining a new complex valued matrix $\xi_{ij}$ as
\begin{equation}
  \zeta_i (t, \vec k) \equiv \xi_{ij} (t, k) \hat a^j (\vec k) + \xi_{ij}^* (t, k) \hat a^{\dagger, j} (- \vec k)\ ,
  \label{eqn:zeta_modes}
\end{equation}
and similarly for ${\delta \rho}_i(t, \vec k)$, which is related to $ {\delta \phi}_i (t, \vec k)$ and its derivatives.
Substituting $q_{ij}(t,k)$ and its derivative into Eq.~\eqref{eq:zeta_spatfloat} gives

\be 
\begin{split}
  \xi_{ij}(t,k) &= \frac{q_{ij}'(t,k)}{3\phi_i'(t)} - \frac{1}{6} \sum_m \phi_m'(t) q_{mj}(t,k)  \\
                 &+ \left[ \frac{V_{,i}(t)}{3 H^2(t) \phi_i'^2(t)} \right] q_{ij}(t,k)\ .
  \label{eqn:xi_expression}
\end{split}
\ee
where $(')$ is a derivative with respect to $e$-folds $N$. A similar expression is available for $\xi_{ij}^* (t, k)$, which is linearly independent of $\xi_{ij}$.
We evaluate this quantity by evolving the $q_{ij}$ (or $\psi_{ij}$) and background quantities numerically as a function of $t$ for a given $k$.\footnote{For more details of the numerical methodology see Ref.~\cite{Price:2014xpa}.}

We expand $\zeta_i(\tosc,{\bf x})$ in terms of field fluctuations at horizon exit,
\be\label{eq:field_fluc_horiz}
\zeta_i(\tosc,{\bf x}) = \sum_{j=1}^\calN C_{ij}\delta\phi_j(t_\star,{\bf x})\ ,
\ee
where $C_{ij}$ is a real matrix.  Substituting our ${\delta \phi}_i(t, \vec k)$ from Eq.~\eqref{eq:modematrix} into Eq.~\eqref{eq:field_fluc_horiz} gives
\begin{equation}
  \xi_{ij}(t,k) = \sum_m C_{im}(t) q_{mj}(t_\star,k)\ .
  \label{eqn:xi_of_c}
\end{equation}
While Eq.~\eqref{eqn:xi_of_c} is not invertible for general $q_{mj}$, we match to the slow-roll approximation above by first discarding the off-diagonal elements of the perturbation matrix $q_{ij}$ at horizon crossing and define a vector $v_j(t_\star,k)$ as
\begin{equation}
  q_{ij}(t_\star,k) \equiv \mathrm{diag}[v_1(t_\star,k), \dots, v_N(t_\star,k)]\ .
  \label{eqn:XXX}
\end{equation}
Since $q_{ij}$ is complex and $C_{ij}$ is real, we take
\begin{equation}
  C_{ij} \approx - \mathrm{sgn} \, (\mathrm{Re}[q_{ij}]) \frac{|\xi_{ij}(t,k)|}{|v_j|}\ ,
  \label{eq:cij_final}
\end{equation}
where the overall sign is chosen to match the two-field results of Ref.~\cite{Meyers:2013gua}.

With this, we have all the ingredients to relate the curvature fluctuations at the end of reheating to the quantum fluctuations during inflation as in Eq.~\eqref{eq:delta_N} where using Eqs.~\eqref{eq:zetaN1}~and~\eqref{eq:field_fluc_horiz}, the derivative of the number of $e$-folds of expansion can now be written as
\begin{equation}
N_{,i}=\sum\limits_jW_jC_{ji}\ .
\end{equation}


\begin{figure*}[t!]
\centering
\includegraphics[width=0.9\textwidth]{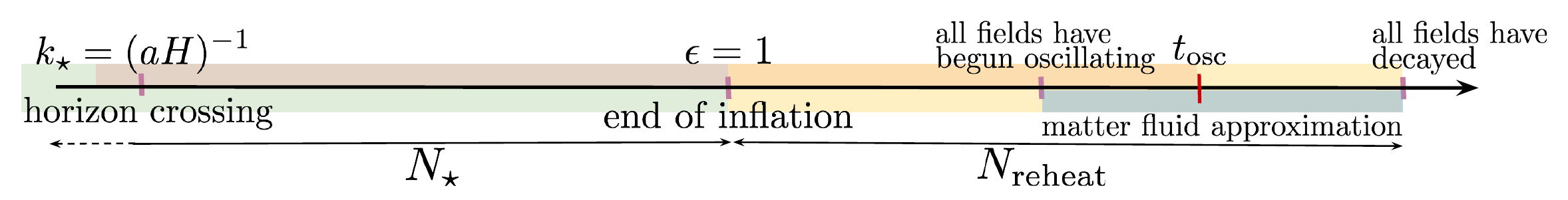}
\caption{Timeline of inflation and reheating.  The method we describe numerically evolves the perturbed scalar field equations until all fields have begun oscillating, after which we switch to evolving unperturbed fluid equations.}\label{fig:timeline}
\end{figure*}


\section{Numerical implementation}\label{sec:numerical}

While the recursive definition of the reheating parameters introduced in Sec.~\ref{sec:sudden_dec} require solving numerically the homogeneous background equations until the end of reheating, it is sufficient to evolve the field fluctuations only until a few $e$-folds into the phase of coherent oscillations, after which the curvature perturbations of the fields $\zeta_i(t^\text{osc}_i,\textbf{x})$ are individually conserved. The prescription for solving such a system of equations will typically involve evolving first the background equations in order to determine the number of $e$-foldings $N_\star$ at which the pivot scale (which we take to be $k_{\textnormal{piv}}=0.05\ \textnormal{Mpc}^{-1}$)  leaves the horizon during inflation, and then the first order fluctuations for each field from deep inside the horizon until the time where the curvature perturbations are conserved. Note that since we mainly want to explore the impact of reheating on inflationary observables, we will sample from many different $\Gamma_i$ distributions while keeping the parameters describing the inflationary model unchanged. Hence this approach is quite inefficient for our purposes, as it requires solving for the inflationary dynamics as well as reheating for each assignment of decay parameters $\Gamma_i$. Instead in this work we have chosen to solve the inflationary fluctuations on a grid of $N_\star$ values in the range $N_\star\in(40-60)$ and perform a local linear fit to determine\footnote{In order to smooth the small round-off error in our simulations, and to capture the underlying scaling with $N_\star$, we linearly fit the $C_{ij}$ matrix elements from the elements of the calculated grid within $\pm1$ $e$-fold of the desired $N_\star$.} the individual elements of the $C_{ij}$ matrix in Eq.~\eqref{eq:cij_final} for a given $N_\star$.

Following the methods outlined in~\textsc{MultiModeCode}~\cite{Mortonson:2010er,Easther:2011yq,Norena:2012rs,Price:2014xpa}, we first solve the Klein-Gordon equations for the homogeneous background fields with initial conditions $\phi_{i,0}$ which determines the field-space positions at the end of the inflation defined by $\epsilon=1$.  We then continue evolving the background fields after the end of inflation, well into the oscillatory phase. For the simulations in this paper, we have evolved the field equations until each field $\phi_i$ has crossed its minimum 5 times, although the exact number does not effect the results significantly after each field has oscillated a few times. Knowing the times for the end of inflation and the onset of coherent oscillations, the ${C}_{ij}$ matrix can be calculated by evolving the mode equations as described in Sec.~\ref{sec:inf_perturbations} for a given value of $N_\star$. We calculate the $C_{ij}$ matrix during the oscillatory period and evaluate the average of the maximum and minimum values for each $C_{ij}$ matrix entry. We use this averaged $C_{ij}$ matrix in calculating the observables. For the reasons explained above, we repeat this step multiple times while varying the quantity $N_\star$.

Since the exact value of $N_\star$ and the normalization of the potential $V_\star$ will depend on the details of reheating, we first solve the post-inflationary dynamics for some fiducial values $V^\text{fid}_\star$ and $N^\text{fid}_\star$. We calculate the $W_i$ array by solving the scalar field equations for the background solution, using the end-of-inflation values $\phi_{i,\mathrm{end}}$ as the new initial conditions. Once a field $\phi_i$ has crossed the minimum of its potential, we turn on the decay term in its equation of motion, which sources the corresponding radiation fluid $\rho^{\gamma}_{i}$ for that field. After all the fields have passed through their potential minima and started decaying into radiation, we stop evolving the Klein-Gordon equations and switch to a fluid description with equations of motion
\bea
	\dot{\rho}_{{i}}+3H{\rho_{i}} &=& -\Gamma_{{i}}{\rho_{i}} \nonumber \\
    \dot{\rho}^\gamma_i + 4H\rho^\gamma_i &=& \Gamma_{i}{\rho_{i}} \, ,
\eea
with the Hubble rate given by the Friedmann equation
\begin{equation}
3H^2=\sum\limits_i(\rho_{i}+\rho_{i}^\gamma)\ .
\end{equation} 
Note that the fluid densities have a mild dependence on when this transition is implemented, but the change to observables is negligible compared to the full range of predictions. We allow this fluid simulation run until all matter fluids have decayed into radiation. From the results of this numerical evolution, we are able to read off the quantities we need to apply the sudden decay approximation and determine the final curvature perturbation in the adiabatic limit at the end of reheating. Each time a decay rate becomes equal to the Hubble rate $\Gamma_i = H$, we evaluate the sudden decay parameters $r_{ij}$ described in Sec.~\ref{sec:sudden_dec}. After all the fields have decayed into radiation, we assume that all decay products quickly come to thermal and chemical equilibrium.  The solutions will then rapidly approach the adiabatic limit, and we can calculate the curvature perturbation and its power spectrum as described in Eq.~\eqref{eq:delta_N} and Eq.~\eqref{eq:curv_power}. This calculation results in a scalar amplitude given by $\mathcal{P}_\zeta^\text{fid}$ which then needs to be rescaled to match observations.    

The amplitude of the scalar fluctuations is fixed by the observations of the CMB anisotropies to be $\mathcal{P}^\text{\tiny{CMB}}_\zeta\approx2.142\times10^{-9}$ \cite{Ade:2015xua}. We rescale the inflationary potential in order to set the power spectrum calculated in Eq.~\eqref{eq:curv_power} equal to this value. The relative quantities transform under the rescaling of the potential as $V\rightarrow \alpha V$ follows: 
\begin{equation}
\rho\rightarrow\alpha\rho\ \ , \ \  H\rightarrow\alpha^{\frac12}H\ \ , \ \  \mathcal{P}_\zeta\rightarrow\alpha\mathcal{P}_\zeta\ \ , \ \ \zeta\rightarrow\alpha^{\frac{1}{2}}\zeta\ , 
\end{equation}
where the scaling for our purposes is $\alpha=\mathcal{P}^\text{\tiny{CMB}}_\zeta/\mathcal{P}_\zeta^\text{fid}$. Having solved the dynamics of reheating, we also know from Eq.~\eqref{eq:matching_efolds} the quantity 
\begin{equation}
\ln\frac{a_\text{reh}}{a_\text{end}}=N_\text{reheat}\ . 
\end{equation}
Rescaling the potential in order to match the CMB observations in turn fixes the remaining quantities in Eq.~\eqref{eq:matching_efolds} where $N_\star$ (for a given $k_\star$) now takes an exact value (see Fig.~\ref{fig:timeline} for a sketch of the timeline). We then fit the $C_{ij}$ matrix elements corresponding to the calculated $N_\star$ from the grid of $C_{ij}$ matrices we already calculated. This rescaling step after solving the dynamics of reheating is repeated for all simulations. Having determined the value $N_\star$, the corresponding $C_{ij}$ matrix and the $W_i$ array, we calculate the power spectrum and the cosmological observables as described in Sec.~\ref{sec:overview}.


\section{A Case Study}\label{sec:case}

We consider inflation with canonical kinetic terms, a minimal coupling to Einstein gravity and $\mathcal{N}$-quadratic potential, 
\be \label{eq:inf_action}
	S = \int \mathrm{d}^4x \sqrt{-g} \left(\frac{R}{2} -  \sum_i\frac{1}{2} g^{\mu\nu}\partial_\mu\phi_i\partial_\nu\phi_i - \sum_i m_i^2\phi_i^2 \right) \, ,
\ee
a model which has been studied extensively elsewhere, \emph{e.g.}~\cite{Liddle:1998jc,Dimopoulos:2005ac,Easther:2005zr,Battefeld:2006sz,Battefeld:2007en,Kim:2007bc,Kim:2006ys,Ellis:2013iea,Bachlechner:2014hsa}.
We study the regime where one (or a few) field(s) dominates the energy density during inflation while the rest remain sub-dominant.  We achieve this by setting the field masses $m_i$ and initial field positions $\phi_{i,0}$ to be distributed linearly in log-space with equal spacing and the same ordering. In this regime the impact of reheating on the inflationary predictions is maximized when the sub-dominant fields get assigned smaller decay parameters, hence scaling like matter for a longer period, dominating the contributions to the curvature perturbation at the end of reheating. In our simulations we kept the ratio between the maximum and minimum masses constant and equal to $m_{\textnormal{max}}/m_{\textnormal{min}}=10^{3}$ and fixed the initial field positions to be in the range $[10^{-3},20]\ \text{M}_\text{pl}$.

We are interested in determining how reheating impacts two-point statistics for a wide range of decay rates, and so we sample the very large parameter space as follows. First, we take the decay rates to be determined by the mass hierarchy as 
\begin{equation}\label{eq:set_ric}
\Gamma_i:= 10^{-4}H_{{\rm end}}\left(\frac{m_i}{m_{\textnormal{max}}}\right)^{\alpha}\ , 
\end{equation}
where $H_{{\rm end}}$ is the Hubble parameter at the end of inflation, for some choice of the parameter $\alpha$. 
Next, we perform a permutation $\sigma_i$ on this first set of decay rates randomly chosen from the $\mathcal{N}!$ possible permutations in order to generate another set $\Gamma_i = \sigma_i(\Gamma)$.   We perform this same procedure for several choices of the parameter $\alpha$ which allows us to adjust the hierarchy between the decay parameters. In all cases, the minimum decay rate is bounded from below by Big Bang nucleosynthesis which constrains the energy scale at the end of reheating to be larger than about 4~MeV, and perhaps higher if fields decay into hadrons~\cite{kawasaki1999cosmological,kawasaki2000mev}, and the maximum decay rate is constrained by the validity of sudden decay approximation to be less than the Hubble parameter at the end of inflation $\Gamma_\text{max}< H_\text{end}$. Note that increasing (decreasing) the value of the maximum decay constant $\Gamma_\text{max}$ will in turn increase (decrease) the value of $N_\star$ that satisfies Eq.~\eqref{eq:matching_efolds}. The results shown in this paper have values of $N_\star$ lower than the instant reheating case, in the range $N_\star\in(45-55)$. 


\begin{figure}[t]
\centering
\hspace*{-0.2cm}\includegraphics[width=0.5\textwidth]{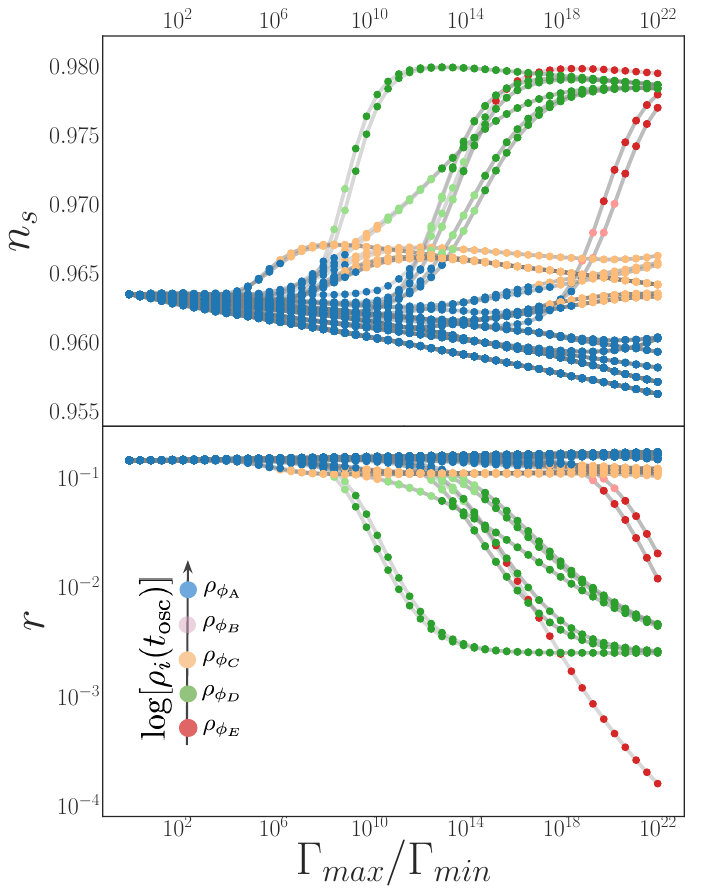}
\caption{
The effect of reheating on the scalar spectral index $n_s$ and tensor-to-scalar ratio $r$  for the case study in Sec.~\ref{sec:case} with $\mathcal{N}=5$ fields with $k_\text{piv}=0.05\ \text{Mpc}^{-1}$. Colored circles show the results from simulations for a particular choice of decay rates $\Gamma_{i}$, chosen as described in text following Eq.~\eqref{eq:set_ric}. Each line (in gray) connects the results from simulations with an identical ordering of $\Gamma_i$ values while the parameter $\alpha$ which determines their spacing is varied in the range $\alpha\in[0,\tiny{\frac{22}{3}}]$ in 50 steps. We plot the results from all possible 120 permutations of $\Gamma_{i}$. The colors mark the field that has the largest $N_{,i}$ at the end of reheating, for a particular simulation (except for the circles with lighter color, which highlight the simulations with the most energetic $\phi_A$ field still having the largest $N_{,i}$, although a second field is within $50\%$ of $N_{,\phi_A}$). The labeling of the fields is ordered with respect to their energy densities at the end of inflation $\rho_{{i}}(t_\mathrm{osc})$ where $\phi_A$ has the largest energy density and $\phi_E$ has the smallest.}\label{fig:obs}
\end{figure}

Fig.~\ref{fig:obs} demonstrates the effect of reheating on the two-point observables, the spectral index $n_s$ and the tensor-to-scalar ratio $r$. The results from numerical simulations described in Sec.~\ref{sec:numerical} are plotted with colored circles. Each line (in gray) connects the results from simulations with the decay rates assigned with the same permutation while the parameter $\alpha$ is varied in the range $\alpha\in[0,\tiny{\frac{22}{3}}]$. The colors of the circles indicate the field with the largest measured $N_{,i}$ for that simulation. As shown in Eq~\eqref{eq:zetaN1}, this parameter depends on two quantities, the $C_{ij}$ matrix and the $W_i$ array which operates on this matrix.

For the $\mathcal{N}$-quadratic case study, $C_{ij}$ matrix has a simple structure where its diagonal elements are significantly larger than its off-diagonal elements. Since fluctuations grow larger in the less massive field directions, the values of the diagonal elements associated with these fields are also larger. Hence, for this study, the subset of simulations where reheating has a significant impact on observables are ones with particular $\Gamma_i$ assignments resulting in the corresponding $W_i$ arrays to preferentially dampen contributions from the heavier fields, while enhancing those from the lighter fields. These simulations are shown with varying colors in Fig.~\ref{fig:obs} where a large impact on observables is obtained when contributions from the lighter fields $\phi_{C,D,E}$ are enhanced.

For most choices of decay rates, the predictions for $n_s$ and $r$ lie very close to the predictions of a model with a single scalar field in a quadratic potential.  The predictions that deviate from this result essentially interpolate between a single field regime and a curvaton-like scenario where a given sub-dominant field dominates the effect on observables, resulting in predictions to asymptotically converge on narrow lines of $n_s$ and $r$ predictions, as can be seen in Fig.~\ref{fig:obs}. The values of the observables corresponding to these lines depend on the masses and values of the fields at horizon crossing, or in other words, on the details of the inflationary model. The total range of predictions in these scenarios therefore depends on the choice of the inflationary model parameters.


\begin{figure*}[t]
\centering
\includegraphics[width=0.7\textwidth]{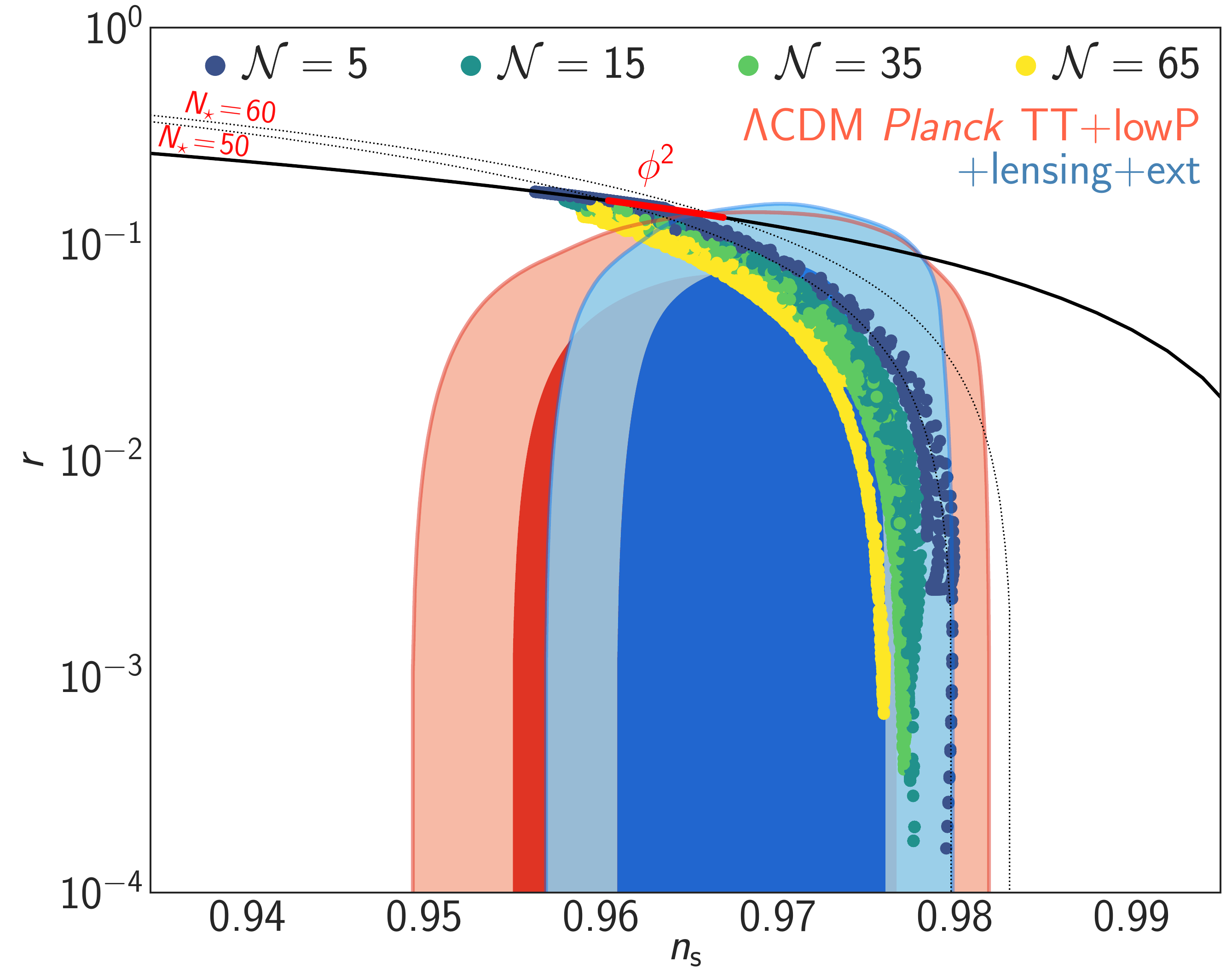}
\caption{The predictions of the $\mathcal{N}$-quadratic inflation case study described in Sec.~\ref{sec:case} for the scalar spectral index $n_s$ and tensor-to-scalar ratio $r$ from the sudden decay approximation,  plotted with the \textit{Planck}~2015 constraints using $k_\text{piv}=0.05\ \text{Mpl}^{-1}$ and assuming zero running.  Dashed lines show the predictions from single-field inflation models with monomial potentials where the pivot scale exits the horizon 50 or 60 $e$-folds before the end of inflation (denoted by $N_\star$). The thick black line is the prediction of single-field quadratic inflation. The colored points are the results from our simulations with $\mathcal{N}=5,\ 15,\ 35\ \text{and}\ 65$ fields. See Sec.~\ref{sec:case} for the details of how the decay rates are chosen. Note that only a small subset of possible choices of the decay rates leads to predictions which differ significantly from the single-field case. In particular, for simulations with a large number of fields $\mathcal{N}\geq15$, only scenarios in which the decay rates share nearly the same hierarchy as the masses lead to predictions with very low $r$. }\label{fig:r_ns}
\end{figure*}

Fig.~\ref{fig:r_ns} summarizes our results for the $\mathcal{N}$-quadratic inflation case study with perturbative reheating and sudden decay approximation. Obtained values for the spectral index $n_s$ and tensor-to-scalar ratio $r$ for simulations with $\mathcal{N}=5,\ 15,\ 35,\ 65$ fields are plotted with the $Planck$~2015 contours \citep{Ade:2015lrj} with the pivot choice $k_\text{piv}=0.05\ \text{Mpc}^{-1}$ and the theory predictions for the single-field quadratic inflationary potential. In populating the $n_s-r$ plane, we show results for the $\Gamma_i$ chosen from all 120 permutations for the models with 5 fields, and for $\mathcal{N}\geq 15$ we show only the results where the $\Gamma_i$ are ordered similarly to the field masses.\footnote{This is a exceedingly small subset of all possible permutations for a model with many fields $\mathcal{N}\gg2$.} The density of points in this figure does not represent a simple measure on the input $m_i$ and $\Gamma_i$ parameter space, but are chosen to highlight the wide range of observable parameter values accessible in these scenarios. Most of the possible permutations, which are \textit{outside} this set, fall near the quadratic inflation predictions (solid black line). 


\section{Discussion}

We have developed a method to treat the impact of reheating on observables following multiple-field inflation.  We have shown how to treat the effects of reheating semi-analytically, greatly reducing the computational cost to make definite predictions with multiple-field models.

Our results focused on one specific form for the inflationary potential, although our method applies much more broadly.  Multiple-field models of inflation have a very rich parameter space which remains largely unexplored.  The techniques described in this work allow for a thorough exploration of this space, including the potentially very important impact of reheating following multiple-field inflation, as has recently been done for a set of two-field models~\cite{Hardwick:2015tma,Vennin:2015egh,Hardwick:2016whe}. We restricted numerical results to $\mathcal{N}$-quadratic inflation with specific choices for both the hierarchy of masses and the initial conditions. We showed that reheating can have an effect on the predictions of multiple-field inflation. For the scenarios we studied, reheating has a significant impact on observables only when the lightest fields are assigned very low decay rates (this is the case that realizes curvaton-like behavior). For choices of parameters where this relation is not present we found almost no sensitivity of the primordial curvature perturbation to the physics of reheating (apart from the dependence on $N_\star$ which is present even in single-field models). At large $\mathcal{N}$ we therefore found only a very small fraction of the tested scenarios exhibited sensitivity to reheating. Different choices of parameters would lead to a different set of perturbations predicted at the end of inflation, and also a different range of predictions for observables following reheating. Our focus has been on exploring a restricted set of initial conditions and model parameters but it would be interesting to perform a statistical analysis of the model as described in \cite{Price:2015qqb}. 

Looking beyond $\mathcal{N}$-quadratic inflation, our method requires only that scalar fields oscillate about quadratic minima, but there is nothing about our technique that restricts the form of the potential away from the minimum, and in fact a straightforward extension of the methods presented here would allow treatment of non-quadratic minima as well.  The effects of reheating are expected to be greater than those shown here for more general choices of potential~\cite{Meyers:2013gua,Vennin:2015egh}.

While the need to include a detailed model of reheating makes multiple-field models of inflation inherently more complicated, the dependence of observables on the reheating phase also presents an opportunity.  Very little is known about how reheating took place, though the sensitivity of observables to reheating following multiple-field inflation may allow more information to be gleaned about this weakly constrained phase of the cosmic history than is possible for single-field models~\cite{Martin:2014rqa,Martin:2014nya,Drewes:2015coa,Martin:2016oyk,Hardwick:2016whe}.

We focused here on the two-point statistics of curvature perturbations, though it would be very interesting to extend our results to include the study of primordial non-Gaussianity~\cite{Bartolo:2004if}.  Unlike single-field inflation models, multiple-field inflation models are capable of producing detectable levels of local-type non-Gaussianity~\cite{Maldacena:2002vr,Creminelli:2004yq}, therefore making calculation of higher-order statistics a natural next step for the tools we have developed here.  Treatment of non-Gaussianity would require carrying out calculations to the next order of perturbation theory, but the general techniques spelled out here should apply without much modification.

Reheating is a necessary component of any successful inflationary model.  For single-field inflation the predictions of observables are sensitive only to the integrated expansion history during reheating.  However, the details of reheating following multiple-field inflation have an important and direct impact on the evolution of cosmological perturbations, and therefore must be treated carefully when predicting the observable outcomes of these models.  We presented here a method to make this treatment tractable.

\acknowledgments

The research leading to these results has received funding from the European Research Council under the European Union’s Seventh Framework Programme (FP/2007–2013) / ERC Grant Agreement No. [308082].
LCP was supported in part by the Department of Energy under grant DESC0011114.  JF acknowledges funding from
the ERC Consolidator Grant STRINGFLATION under the HORIZON 2020 contract
no. 647995.
J.M. was supported by the Vincent and Beatrice Tremaine Fellowship. SCH is funded by the Imperial College President's
Scholarship. SCH would like to thank Imperial College High Performance Computing Service at Imperial College London (UK) for providing computational resources.

\bibliographystyle{JHEP}
\bibliography{bib_file}

\end{document}